\documentclass[aps,pre,11pt,soul,floatfix,tightenlines,onecolumn,superscriptaddress,final]{revtex4-2}

\usepackage[normalem]{ulem}
\usepackage{amsfonts}
\usepackage{color}
\usepackage{epsfig}
\usepackage[breaklinks,colorlinks = true,linkcolor = blue,urlcolor=blue,citecolor=blue]{hyperref}
\usepackage{subcaption}
\usepackage{ragged2e}
\DeclareCaptionJustification{justified}{\justifying}
\usepackage{subfiles}
\usepackage{graphicx}
\usepackage{dcolumn}
\usepackage[nointegrals]{wasysym}
\usepackage{dsfont}
\usepackage{bm}
\usepackage{hyperref}
\hypersetup{
    colorlinks=true,
    linkcolor=blue,
    citecolor=teal,
    urlcolor=blue
}
\usepackage{amsmath,amssymb,enumitem}

\usepackage{pifont}

\usepackage{cancel}

\usepackage[none]{hyphenat}
\usepackage{mathrsfs}
\usepackage[inline]{trackchanges}

\newcommand{\mc}[1]{\mathcal{#1}}

\newcommand{\mb}[1]{\mathbb{#1}}

\newcommand{\grad}[1]{\vec{\nabla}#1}

\newcommand{\mean}[1]{\langle #1 \rangle}
\DeclareRobustCommand{\eneg}{\hat{e}_{-}}
\DeclareRobustCommand{\epos}{\hat{e}_+}

\renewcommand{\d}[3]{\frac{d^{#3} #1}{d #2^{#3}}}

\renewcommand{\vec}[1]{\boldsymbol{#1}}

\renewcommand{\div}[1]{\vec{\nabla} \cdot {#1}}

\usepackage[dvipsnames]{xcolor}
\usepackage{xparse,wrapfig}

\makeatletter

\DeclareRobustCommand{\rc}[1]{\@ifclasswith{revtex4-2}{final}{#1}{\textcolor{olive}{#1}}}

\DeclareRobustCommand{\rg}[1]{\@ifclasswith{revtex4-2}{final}{#1}{\textcolor{magenta}{#1}}}
\DeclareRobustCommand{\rcomm}[1]{\@ifclasswith{revtex4-2}{final}{}{\textcolor{OliveGreen}{[#1]}}}

\newcommand{\eref}[1]{Eq.~\eqref{#1}}
\newcommand{\fref}[1]{Fig.~\ref{#1}}
\makeatother

\DeclareMathOperator{\Tr}{Tr}

\usepackage[mathlines]{lineno}
\usepackage{showlabels}
\usepackage{pgf,import,lmodern,microtype}
\usepackage[export]{adjustbox}  
\usepackage{subcaption,tikz,setspace}
\usepackage[utf8]{inputenc}        
\newcommand{\cs}{C~}
\newcommand{\es}{A~}
\newcommand{\su}{S~}
\usepackage{parskip,float,orcidlink}
\begin{document}
\title{Caustics of finitely dense inertial particles} 

\author{C. Rajarshi \orcidlink{0009-0009-1747-3916}}
\email{rajarshic.edu@gmail.com}
\affiliation{International Centre for Theoretical Sciences, Tata Institute of Fundamental Research, Bangalore 560089, India}
\author{Rama Govindarajan \orcidlink{0009-0007-4794-4964}}%
\email{rama@icts.res.in}
\affiliation{International Centre for Theoretical Sciences, Tata Institute of Fundamental Research, Bangalore 560089, India}

\justifying
\begin{abstract}
Estimating collision rates is of immense importance in particle-laden flows. An economical way of doing this is to directly identify incidences of caustics, or extreme clustering, by tracking particle velocity gradients in the neighborhoods of individual particles. The objective of this work is two-fold. (i) We find conditions under which caustics form, in point-vortex flow and in two-dimensional turbulence. While caustics are known to form in regions of strain, we show that the \rc{velocity alignment with strain directions} is key. Particles must remain in compressional strain throughout the process to form caustics, whereas survivor particles: which visit high strain but do not form caustics, briefly go through extensional strain during the early part of the process. This enables survivor particles to attain significantly straighter paths, and to move faster, whereas caustics particles follow paths of high curvature and move slower. As a result, caustics particles stay longer in high-strain regions than survivors. (ii) We ask about the effect of finite particle density, where the particle is denser than the background fluid. We show that finite-density particles need to sample stronger background strain than infinite-density ones to trigger caustics, but our other findings are universal across particle density.

\end{abstract}

\maketitle

\section{Introduction} 
Pollens that spread forests, pollutants that smother cities, water droplets that rain, and oceanic plankton that control the global carbon cycle are all particles suspended in, and heavily dominated by, a turbulent background flow. Such particles are often denser than the background fluid and are of finite size. They thus have non-zero inertia and respond to turbulence by centrifuging out of vortical regions, accumulating in high-strain regions. This phenomenon is called preferential clustering \cite{crowe1985,balkovsky2001,calzavarini2008a}. Particles undergoing extreme clustering collide with each other, forming caustics. During caustics, particles distributed initially in two or three dimensions come so close together that they may occupy a smaller spatial dimension. These collisions or caustics are central to rain formation \cite{falkovich2002,shaw2003,wilkinson2006,pumir2016}, planetesimal formation in protoplanetary disks \cite{guttler2010,schrapler2011,wada2013,birnstiel2024}, and for maintaining populations of small creatures in the ocean by enhancing reproduction rates.

Though rare in dilute suspensions, caustics are thus crucial for our very survival. The mechanism of how turbulence promotes caustics is not entirely clear, and this paper deals with one aspect: we ask how high a strain a finitely dense caustics particle needs to experience, and what makes other particles that do experience such strain not form caustics.
The situations of our interest involve $O(10^j)$ particles, where $j$ is far too large for numerical solutions of individual particle dynamics to be feasible. One may resort to continuum descriptions where the particles are described by a compressible velocity field $\vec{v}(t,\vec{x})$ on the space $\vec{x}$. Caustics are singular points of extreme compression of this field, where the rate of change of the particle number density diverges at finite time, i.e., $\nabla \cdot \vec{v} \to -\infty$ \cite{falkovich2002}. 
These can be identified by evolving the dynamics of representative particles as well as particle velocity gradients at each particle's location, directly in a Lagrangian sense. This captures the behavior of an infinitesimal particle parcel in its neighborhood \cite{falkovich2002,meibohm2021}. With this approach, we may characterize caustics formation in a turbulence field by following a relatively small number of particles.

 Caustics have been identified \cite{falkovich2002,shaw2003,wilkinson2006} as a potential mechanism for abrupt droplet growth causing rain showers. \citet{falkovich2007} found collision rates in cloud droplets due to caustics to be higher than previous geometric arguments. Later \citet{bhatnagar2022} looked for, but did not find, a scaling for the rate of caustics formation of infinitely dense particles as a function of particle size or Stokes number. Modeling the fluid velocity gradient as white noise, \citet{barta2022} found that there are infinite ways to form caustics in the phase space of invariants of the particle velocity gradient tensor. We note, however, that velocity gradients in turbulence are highly non-Gaussian \cite{meneveau2011}. Studying collisions between individual infinitely-dense particles, \citet{picardo2019} showed that high-strain regions, just outside vortices, facilitate head-on collisions. 

The present study builds on two others. The first is the work of \citet{meibohm2021}, who derive a caustics condition, expressed as an inequality, in two-dimensional flow for small infinitely-dense particles, and subsequently \cite{meibohm2023} a condition in three-dimensional flow. As we discuss later, these conditions are derived for particles `frozen' in place. Direct numerical simulations (DNS) \cite{meibohm2024}, showed that caustics events do not always satisfy these inequalities. The second is that of \citet{batge2023}, who found the inequality to not be a sufficient condition for infinitely-dense particle caustics, and assuming that the fluid and the particle gradient matrices commute with each other during caustics, showed analytically that particles need to remain within regions of extreme strain for long enough times in order to undergo caustics. They further confirmed this through direct numerical simulations.

While caustics are known to form in strain regions, we show how {the kind of strain} affects the outcome. Secondly, all real particles are finitely dense, and their caustics formation has not been studied, to our knowledge. We {study caustics } in two-dimensional flows for particles denser than the fluid. We extend the Lagrangian equation for detecting caustics to finite density particles in \S \ref{sec: system}, and discuss its advantages. Next in \S \ref{sec:caus_cond}, we follow \cite{meibohm2021,meibohm2023} to study caustics formation of artificially frozen particles, since it is analytically appealing. Thereafter, we study caustics formation of particles moving as they should, in two flows: a toy flow in \S \ref{pv} consisting of a single point-vortex, and two-dimensional turbulence in \S \ref{turb}. Lastly, in \S \ref{conc}, we summarize our findings and discuss their implications.

\section{Setting up of the problem}\label{sec: system}
We assume one-way coupling, i.e., that particles are too small to affect the flow and are in dilute enough suspension so as not to influence each other. In the limit of negligible Reynolds number, a spherical particle of velocity $\vec{v}(t)$ and position $\vec{x}_p(t)$ at time $t$ obeys the Gatignol-Maxey-Riley equation \cite{maxey1983,gatignol1983}, which, in its simplest form, reads, in non-dimensional terms, as
    \begin{align}
    \label{NDeom}
    \begin{split}
      \d{\vec{x}_p}{t}{} = \vec{v} \quad;\quad\d{\vec{v}}{t}{} = \frac{\alpha}{St} \left( \vec{u} - \vec{v}\right) +3\left( 1 - \alpha \right) \frac{D\, \vec{u}}{Dt},
    \end{split}
    \end{align}
  where $\vec{u} = \vec{u}(t,\vec{x}_p(t))$ is the velocity of the background fluid at the particle's position. We have ignored the effects of external forces such as gravity on the particles, and used characteristic flow length and velocity scales $L$, and $L/\tau_f$ respectively, where $\tau_f$ is the evolution timescale of the background fluid. The nondimensional parameters in the problem are the density ratio $\rho \equiv \rho_p/\rho_f$, appearing in terms of the density parameter $\alpha \equiv 2\rho/\left( 2\rho + 1 \right)$, and the Stokes number $St \equiv \tau_p/\tau_f$ where
  $\tau_p = 2 a^2 \rho/(9\nu)$. The subscripts $p$ and $f$ stand for particle and fluid, respectively, $a$ is a particle's radius, and $\nu$ the kinematic viscosity of the fluid. We are interested in particles denser than the fluid, with $2/3 < \alpha \le 1$, where $2/3$ corresponds to neutrally buoyant particles and $1$ to infinitely-dense particles. We note that when $\alpha = 1$, $\vec{v} = \vec{u}$ is a solution of \eref{NDeom}, whereas this is not the case for finitely-dense particles. Added mass and fluid acceleration make up the last term in \eref{NDeom}, which are absent in the dynamics of an infinitely-dense particle. Due to these additional forces, centrifugation by vortices, as well as clustering rates in strain regions 
  are weaker at lower particle densities   \cite{balkovsky2001,calzavarini2008a,volk2008,qureshi2008,calzavarini2008a, fiabane2012,karchniwy2019, petersen2019, motoori2023}. This hints at a lower formation rate of caustics for finitely dense particles. \eref{NDeom} ignores the finite size, lift, and Basset-Boussinesq history effects on the particle. Finite size decreases acceleration fluctuations, as seen for neutrally buoyant particles \cite{calzavarini2009}, and ignoring them restricts us to small particle sizes. Lift forces, proportional to the vorticity the particles experience \cite{mathai2020}, govern the dynamics of bubbles  \cite{mazzitelli2003} but are relatively weak for particles denser than the fluid as they centrifuge out of vortices and live in strain regions. 
  History forces affect clustering at O$(1)$ Stokes number \cite{olivieri2014, daitche2015}. Practical situations concerning caustics and collisions involve small $St$ particles, where \citet{ferry2001} showed that the history force is less important than the added mass. Our recent work \cite{kapoor2024} shows that history effects are quantitative rather than qualitative in the sense that particles of large Stokes number with history forces included behaved like those at smaller Stokes number where history was neglected. Still, whether neglecting Basset-Boussinesq history is justified will be a factor for future evaluation. Given the heavy computational load it entails, it is common to neglect it in early studies of a phenomenon, as we do here. We believe that the physics we uncover will hold with history effects as well.  

\eref{NDeom}, with $d/dt \equiv \partial/\partial t + \vec{v}\cdot \grad{}$, applies to a continuum of particles, under the assumption that $\vec{v}$ is continuous in space. 
The particle continuity equation reads
    \begin{align}
    \begin{split}
        \d{n}{t}{} = - n \div{\vec{v}}.
    \end{split}    
    \end{align}
   We see that at caustics, when $\div{\vec{v}} \to -\infty$, the particle number density $n$ also diverges. In terms of the particle velocity gradient matrix $\mb{Z}  \equiv St\grad{\vec{v}}$ \cite{falkovich2002,meibohm2021}, caustics happen when $\Tr(\mb{Z}) \to -\infty$. From \eref{NDeom}, we can show that for finitely-dense particles, $\mb{Z}$ is related to the fluid velocity gradient matrix $\mb{A} = St\grad{\vec{u}}$ by
    \begin{align}
    \label{Zeom}
    \begin{split}
      \d{\mb{Z}}{t}{}    &= - \frac{\alpha}{St}\left( \mb{Z} -   \mb{A} \right)- \frac{1}{St}\mb{Z}^2+  3(1-\alpha)\left[ \,\frac{ D\, \mb{A}}{Dt}+ \frac{1}{St}\mb{A}^2  \right] \ .
    \end{split}
    \end{align}
   In the limit of infinite particle density, i.e., $\alpha=1$, the above equation reduces to 
   \begin{equation}    
   St~ \frac{d\mb{Z}}{dt} = \mb{A} - \mb{Z} - \mb{Z}^2,
   \label{theirs}
   \end{equation} 
   which has appeared before in the literature (see e.g. \cite{falkovich2002,wilkinson2006,meibohm2023}). For tracking of caustics, \eref{Zeom} presents us with tremendous advantages over both solving for individual particles using \eref{NDeom} and solving for a continuous $\vec v$ field on Eulerian grid points. To obtain reliable statistics out of chance collisions between particle pairs, a numerical approach would need a prohibitively large number of particles. On the other hand, once a single collision occurs in an Eulerian grid, i.e.,  $-\nabla \cdot \vec v$ diverges somewhere in the domain, the simulation cannot be continued, and there is no future after caustics. In the present approach we seed the flow with a finite number of particles and solve for the particle dynamics using \eref{NDeom}, alongside solving for the particle velocity gradient tensor $\mb Z$ using \eref{Zeom}. The singularities now remain restricted to individual particles, and can be artificially removed \cite{falkovich2007}. A noteworthy limitation of this approach is that it cannot capture collisions of faraway particles approaching each other at high speed. 
   
In the limit $St \to 0$, $\vec{v} \to \vec{u}$ by \eref{NDeom},  but by \eref{Zeom}, $\mb{Z}$ does not relax to $\mb{A}$. Thus, clusters once formed cannot be unclustered with ease by the background flow. Also, if the magnitude of the entries of $\mb{Z}$ become large enough that $-\mb{Z}^2$ eclipses all other terms on the right-hand side of \eref{Zeom}, caustics become inevitable. Denoting such a state by $\mb{Z} = \mb{Z}_b$ at time $t=t_b$, caustics occur in finite time {$t_c$ with $(t_c -t_b) \sim -St/\mb{Z}_b$}. This hints at a universality in caustics formation for particles of any density.

\section{Caustics on ``frozen" particles}
\label{sec:caus_cond}

\begin{figure}
\includegraphics[width = \linewidth]{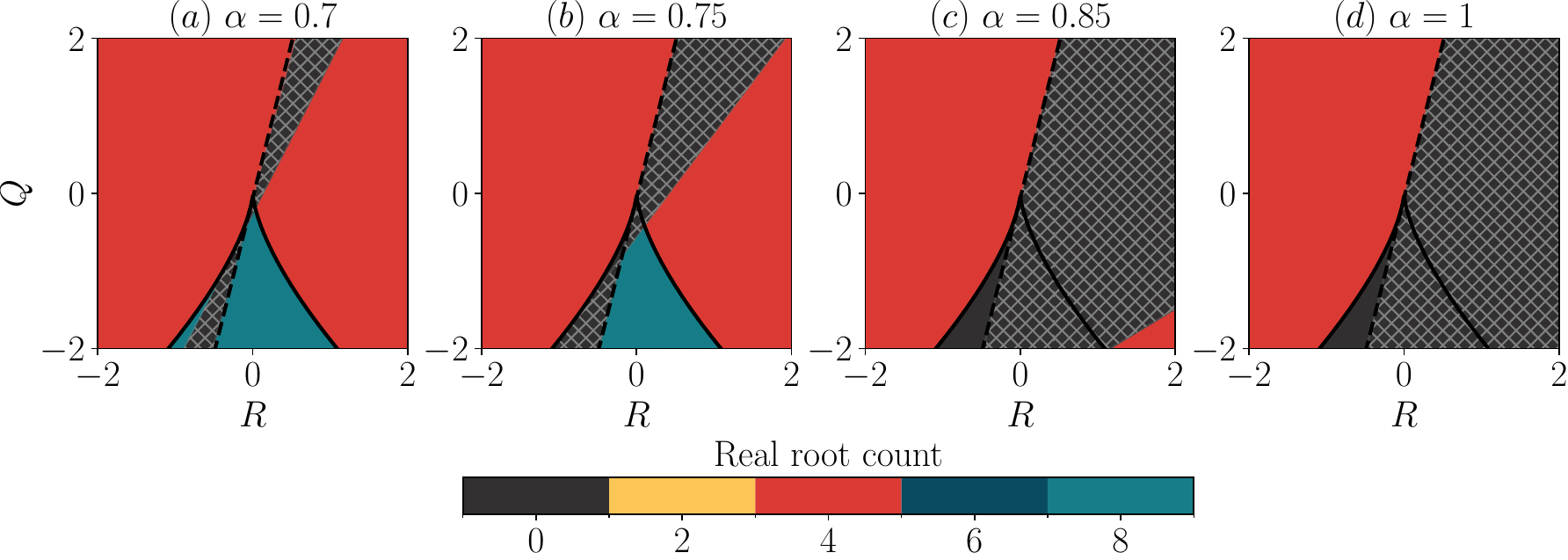}
\caption{Number of real roots of \eref{root-eq} in the $QR$ plane. The hatched area satisfies \eref{caus_cond}, which is a sufficient condition for caustics, while there are some regions with zero roots even outside it. The solid black line is the Vieillefosse line following $Q^3 + 27/4 R^2 = 0$ (see \cite{chong1990} for details). The black dashed lines show $Q - 4R = -1/16$, the caustics condition for $\alpha = 1$.}
\label{inequality}
\end{figure}

Setting $\mb A$ to a constant in \eref{Zeom} has the effect of `freezing' the particles in place in a steady flow, while allowing for local clustering, denoted by $\mb Z$, to evolve. {This is an artificial construct which has its limitations. Still, it is appealing since it enables analytical reduction. We include this section primarily because this method was used previously in literature \cite{meibohm2021,meibohm2023,meibohm2024} for infinitely dense particles.} \citet{meibohm2021,meibohm2023} froze their $\alpha=1$ particles (without explicitly mentioning this), and showed that the system in \eref{theirs} has two options. If a real $\mb{Z}=\mb{Z}_0$ satisfies \eref{theirs} the system may attain this fixed point. If no such $\mb{Z}_0$  exists, and $St \to 0$, caustics inevitably result. Caustics occur whenever $Q - 4R < -1/16$ where $Q = -\Tr(\mb{A}^2)/2$ and $R = -\det(\mb{A})$. In two dimensions (2D), $R=0$, and this reduces to $Q < -1/16$. Note that $Q$ is the Okubo-Weiss parameter that indicates how vortical ($Q>0$) or strain ($Q<0$) dominated a given point in the flow is. We shall see that particles visiting high negative values of $Q$, i.e., high-strain regions, are essential to caustics formation. Following \cite{meibohm2021}, we derive a caustics condition for finitely-dense frozen particles with $St \to 0$:
\begin{align}
\label{neq}
\begin{split}
  -\alpha\left( \mb{Z}- \mb{A} \right) -\mb{Z}^2  + 3(1 - \alpha) \mb{A}^2 \neq 0.
\end{split}
\end{align}
As derived in Appendix \ref{caus-reg-drv}, and defining $\xi \equiv 1-\alpha$, regions of the background flow satisfying 
\begin{align}
\label{caus_cond}
\begin{split}
  \alpha ^2 \left[\alpha ^4+144 \xi^2 Q^2+8 (3 \alpha -1) \alpha ^2 Q\right]-16 \alpha  R \left[\alpha ^2 (9 \alpha -5)+36 \xi^2 Q\right]
     + 1728 \xi^3 R^2 < 0
\end{split}
\end{align}
satisfy the inequality \eqref{neq} for any real $\mb{Z}$, resulting in frozen-particle caustics. For $\alpha = 1$, \eref{caus_cond} reduces to the linear relationship of \cite{meibohm2023}. 
The gray regions with crossed white hatches in \fref{inequality} satisfy \eref{caus_cond}, whereas solid colors correspond to different numbers of real roots of
\begin{align}
  \label{root-eq}
\begin{split}
    -\alpha\left( \mb{Z}- \mb{A} \right) -\mb{Z}^2  + 3(1 - \alpha) \mb{A}^2 = 0,
\end{split}
\end{align}
where caustics may not occur.
In these regions, we find that there is at least one stable root (not shown). We note that \eref{caus_cond} presents a sufficient condition for caustics, and that there can be regions outside it too with zero roots, as seen in \fref{inequality}(c,d), and where caustics can therefore occur. Defining $\delta \equiv \Tr(\mb{Z}) = St \nabla \cdot \vec{v}$ and taking the trace of \eref{Zeom}, we get 
\begin{align}
  \label{div_v_eq}
  \begin{split}
    St\d{\delta}{t}{} = -\alpha\delta - \Tr(\mb{Z}^2) - 6(1 -\alpha) Q.
  \end{split}
\end{align}
We see that for finitely-dense particles, a large negative $Q$ makes a positive contribution to $\delta$, taking it away from caustics. This is remarkable, because high strain is normally expected to be positively correlated with caustics formation. So finite-density particles contain physics competing with the standard clustering effect of strain. 

\section{Caustics in two-dimensional flows}\label{caus-num}

\subsection{Point Vortex Flow}\label{pv}
Before moving to 2D turbulence, it is insightful to study caustics in the vicinity of a point vortex. 
The dimensional velocity field is given by $\vec{u}^* = \hat{\theta} ~\Gamma /2\pi r^*$, where $r^*$ is the dimensional length from the origin, and $\Gamma$ is the circulation. Choosing the length scale $L  = \sqrt{\Gamma \tau_p/2\pi}$ \cite{ravichandran2015}  and timescale $\tau_p$, the nondimensional velocity and the invariant $Q$ become
\begin{equation}
    \vec{u} = \frac{1}{r}\hat{\theta}, \quad {\rm and} \quad Q = -\frac{1}{r^4},
\end{equation} 
whereas the invariant $R=0$ in a 2D flow. The entire flow is thus in a strain field which is singular at the point vortex and rapidly decays away from it.
Using \eref{NDeom}, the non-dimensional equation of motion of particles around a point vortex in polar coordinates $\vec{x}_p = (r,\theta)$ reads
\begin{align}
  \label{nd_ptvrtx}
  \begin{split}
      \ddot{r} + \alpha \dot{r} &= \frac{\mathscr{L}^2 - 3(1-\alpha)}{r^3} \quad 
 ; \quad \dot{\mathscr{L}} = \alpha(1 - \mathscr{L}) \ , 
  \end{split}
\end{align}
 which is an extension to $\alpha \ne 1$ of the expression of \cite{ravichandran2015}, and where $\mathscr{L} = r^2 \dot{\theta}$ is the angular momentum per unit mass, and $\dot{()} = d()/dt$. Our choice of scales renders \eref{nd_ptvrtx} parameter-free, so a given solution holds for particles of any Stokes number when appropriately scaled. We show results for $St=1$.

 \eref{nd_ptvrtx} may be solved for $\mathscr{L}$ to give $\mathscr{L}(t) = 1 - (1 - \mathscr{L}(0))e^{-\alpha t}$. In fact, if we start from $\vec{v}(0) = \vec{u}(r(0))$, we have $\mathscr{L} = 1$ for all time. Caustics happen when two rings of particles at different initial radii cross each other. Due to the circular symmetry of the system, we need to evolve just one particle for each ring. Infinitely-dense particles lying within a threshold radius from the point vortex all form caustics \cite{ravichandran2022}, while those beyond never do. The threshold, however, is only significant in real turbulence where the vorticity is comparable to or greater than the particle's relaxation rate \cite{ling2013}. \citet{ravichandran2015} recognized that the early-time dynamics of particles lying close to the vortex, i.e., in higher strain regions, determine their caustics formation. In other words, caustics is governed by the inner solution in a singular value problem, for which we define the inner variables $ \mc{R}= r/\ell $ and $T = t/\tau$, with $\tau, \ell \ll 1$. 
We follow the usual procedure in singular perturbation problems: of expanding the inner variables in powers of the small parameter $\tau$, and deriving the lowest order equation. The scaling of $l \sim \sqrt\tau$ \cite{deepu2017} yields physical solutions, and our choice of 
$\ell^2 = \tau\sqrt{3\alpha -2}$ yields the following parameter-free equation, valid for all Stokes numbers and all particle densities at the lowest order:
\begin{equation}
   \mc{R}'' = \mc{R}^{-3}, 
\quad {\rm whose \ solution \ is} \quad
\mc{R}(T) = \sqrt{\frac{T^2}{\mc{R}_0^2}+\mc{R}_0^2} \ , 
      \end{equation}
  where $()' = d()/dT$, and $\mc{R}_0 = \mc{R}(0)$. This inner solution is valid close to the vortex origin and always yields caustics. Two rings with initial radius $\mc{R}_0$ and $\mc{R}_0 + \epsilon$ will form caustics at time $T_c$, given by 
  \begin{align}
  \begin{split}
      \sqrt{\frac{T_c^2}{\mc{R}_0^2}+ \mc{R}_0^2} = \sqrt{\frac{T_c^2}{(\mc{R}_0 +\epsilon)^2}+ (\mc{R}_0 +\epsilon)^2} \ .
  \end{split}
  \end{align}
  In the limit $\epsilon \to 0$, a Taylor expansion yields $T_c = \mc{R}_0^2$  and the caustics radius $\mc{R}_c = \mc{R}(T_c)= \sqrt{2}\mc{R}_0$. In the original variables of $t = \tau T$ and $r = \ell \mc{R}$, the caustics radius and time read
  \begin{align}
  \label{inner_caus}
  \begin{split}
      r_c = \sqrt{2}r_0  \qquad ; \qquad t_c = \frac{r_0^2}{\sqrt{3\alpha - 2}} \ .
  \end{split}
  \end{align}  
  \begin{figure}
  \includegraphics[width = 0.8\linewidth]{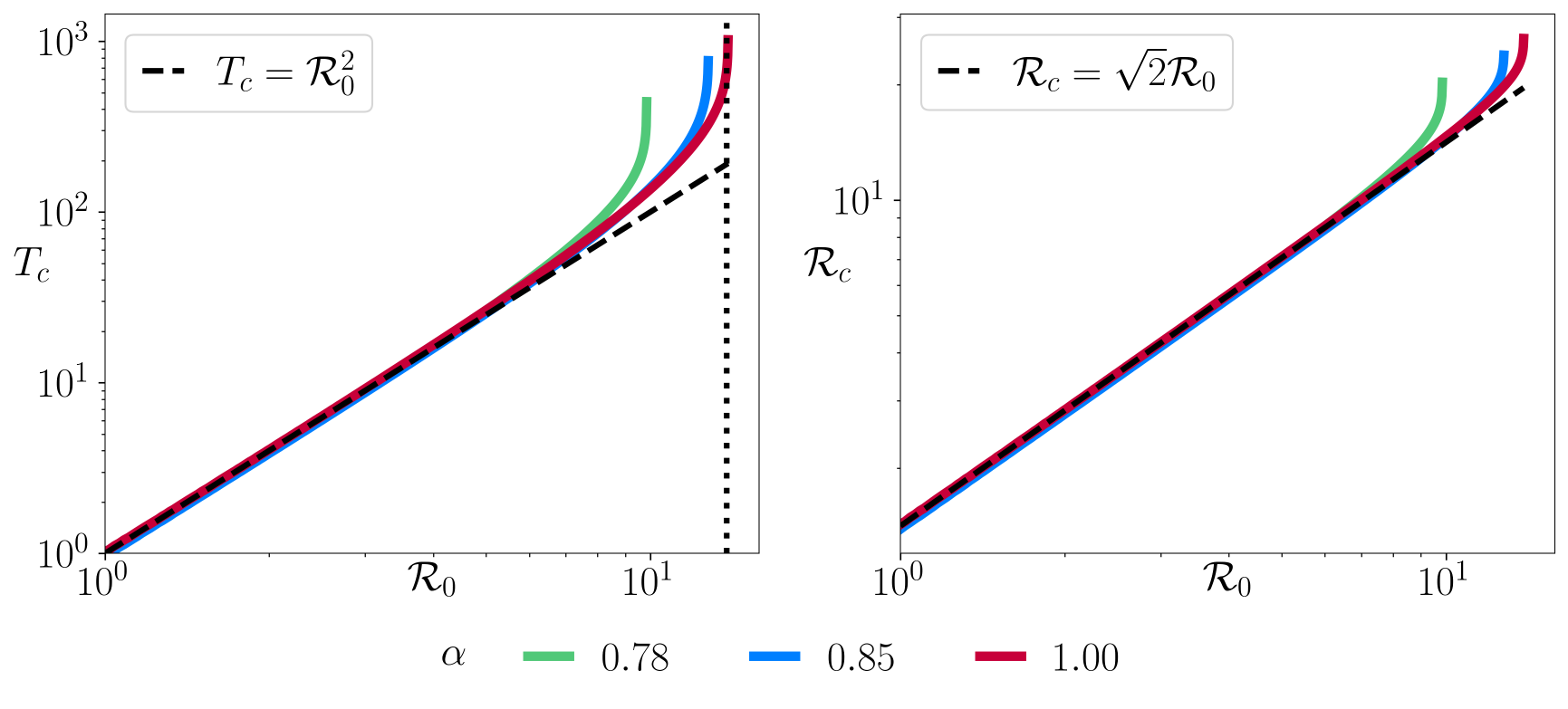}
  \caption{Caustics time (a) and radius (b) in point-vortex flow as a function of the initial radial location, for particles of different density and of any Stokes number. The dashed line shows the inner solutions $T_c = \mc{R}_0^2$, and $\mc{R}_c = \sqrt2\mc{R}_0$. In this scaling, caustics formation in the inner region is independent of particle density. The vertical dotted line indicates the location at which caustics time diverges for $\alpha=1$, and beyond which caustics are not possible.}
  \label{inner-sol}
  \end{figure}
This immediately tells us that the caustics time scales as $(3\alpha - 2)^{-1/2}$, and diverges for neutrally buoyant ($\alpha = 2/3$) particles, ensuring no caustics for them. And the size $\ell$ of the inner region, which is proportional to $r_c$, scales as $(3\alpha - 2)^{1/4}$, shrinking to $0$ for neutrally buoyant particles.
  The scaled caustics radius ($\mc{R}_c$) and time ($T_c$) obtained from numerical simulations of the complete equation \eref{nd_ptvrtx} are shown in colors in \fref{inner-sol}(a,b) respectively. We see that the analytically obtained solution (black dashed lines) matches well with the numerical results close to the vortex origin. When $\mc{R}_0 \gg 1$, the inner solution is no longer valid, and the numerical solution differs from it. Importantly, the caustics time diverges beyond a critical initial radius $\mc{R}_{c0}$ for each $\alpha$. Particles of any density beginning life within their respective $\mc{R}_{c0}$ always form caustics. We call them `C' particles. In contrast, all those starting out further away from the vortex never experience high strain and avoid caustics. Accordingly, we call them `avoiders' or `A' particles. 
  
  We may also identify the region for frozen-particle caustics by solving for $r$ in \eref{caus_cond}:
  \begin{align}
  \label{caustics-sol}
  \begin{split}
    \left[ 4 (3 \alpha -1) - 8 \sqrt{3 \alpha -2}\right]^{\frac{1}{4}} < r\sqrt{\alpha} < \left[ 4 (3 \alpha -1) + 8 \sqrt{3 \alpha -2}\right]^{\frac{1}{4}}.
  \end{split}
  \end{align}
  This result is very different from our moving particle analysis. The reason for this major discrepancy is that \eref{caustics-sol} assumes that a particle that is initially within the region described by \eref{caustics-sol} stays within it until caustics form, {which is an artifical construct}. Moving particles that escape this region in {nondimensional} time shorter than $St$ ({at times shorter than particle response time}) escape caustics even though they started life in a conducive region for frozen-particle caustics. {Thus, although the frozen particle condition is mathematically appealing, it is not very insightful or useful. And, its applicability for finite Stokes particles is greatly limited.}
  
\subsection{Two-dimensional turbulence}\label{turb}

Armed with the insights from point-vortex flow, we investigate caustics of finitely-dense particles in steady state two-dimensional turbulence, by evolving an incompressible turbulent flow field $\vec{u} = (u_x,u_y)$ with vorticity $\omega=\partial_x u_y - \partial_y u_x$, which is evolved according to
\begin{align}
  \label{2DNS}
  \begin{split}
      \partial_t\omega + \vec{u}\cdot\vec{\nabla} \omega = \nu\nabla^2\omega - \gamma \omega  +  f_\omega.
  \end{split}
\end{align}
A small linear drag with coefficient $\gamma$ (here set at $10^{-2}$) is added to prevent a numerical blow-up caused by the inverse energy cascade in $2$D turbulence. Following \cite{pandey2019}, we set the external forcing function $f_\omega = -\nu k_f(\cos(k_f x) + \sin(k_f x))$, where $k_f = 4$ is the forcing wavenumber. The initial vorticity is $\omega(0) = k_f\cos(k_f x)$, and the viscosity $\nu = 8 \times 10^{-6}$. We evolve the flow using a pseudospectral simulation on a $1024\times 1024$ grid with doubly periodic boundary conditions, and a fourth-order Runge-Kutta time-stepping. At the steady state, the Kolmogorov time is $\tau_f = 4.0$ in simulation units. $3.14 \times 10^{5}$ inertial particles of $St=0.225$, unless otherwise specified, that obey \eref{NDeom} are inserted randomly across the flow once the turbulence has reached a stationary state. At this initial time, $\vec v$ and $\mb{Z}$ of each particle are set equal to $\vec u$ and $\mb{A}$, respectively, interpolated to its position. We use fifth-order B-spline interpolation \cite{vanhinsberg2012}, and evolve $\mb{Z}$ according to \eref{Zeom}. {When $\delta = \Tr(\mb{Z})$ reaches large enough negative values, around $\text{O}(-10)$ in our case, $\mb{Z}^2$ dominates all other terms in Eq. \eqref{Zeom}, making caustics inevitable. We set $t_c$ as the time where $\delta$ goes beyond $-20$. We have checked for many cases that when $\delta<-10$, it diverges as $\sim 1/(t-t_c)$ and caustics follow in each case, and that the error we make at setting the caustics time $t_c$ as the time when $\delta = -20$ is minimal due to the rapidity with which $\delta$ diverges. }

To keep the particle number constant beyond caustics, we reinitialize the particle by setting $\mb{Z} \to -\mb{Z}$. This transformation is equivalent to the infinitesimal particle volume element reemerging after caustics as ghost particles \cite{falkovich2007} occupying a small finite volume. Simulations for different $\alpha$ are run up to a time of $125\tau_\eta$, and statistics are collected over this time. {In a separate simulation, by not reinitializing the particles, we have checked that attaining $\delta = -20$ is inevitably and immediately followed by caustics. }

\begin{figure}
\centering
\includegraphics[width = \linewidth]{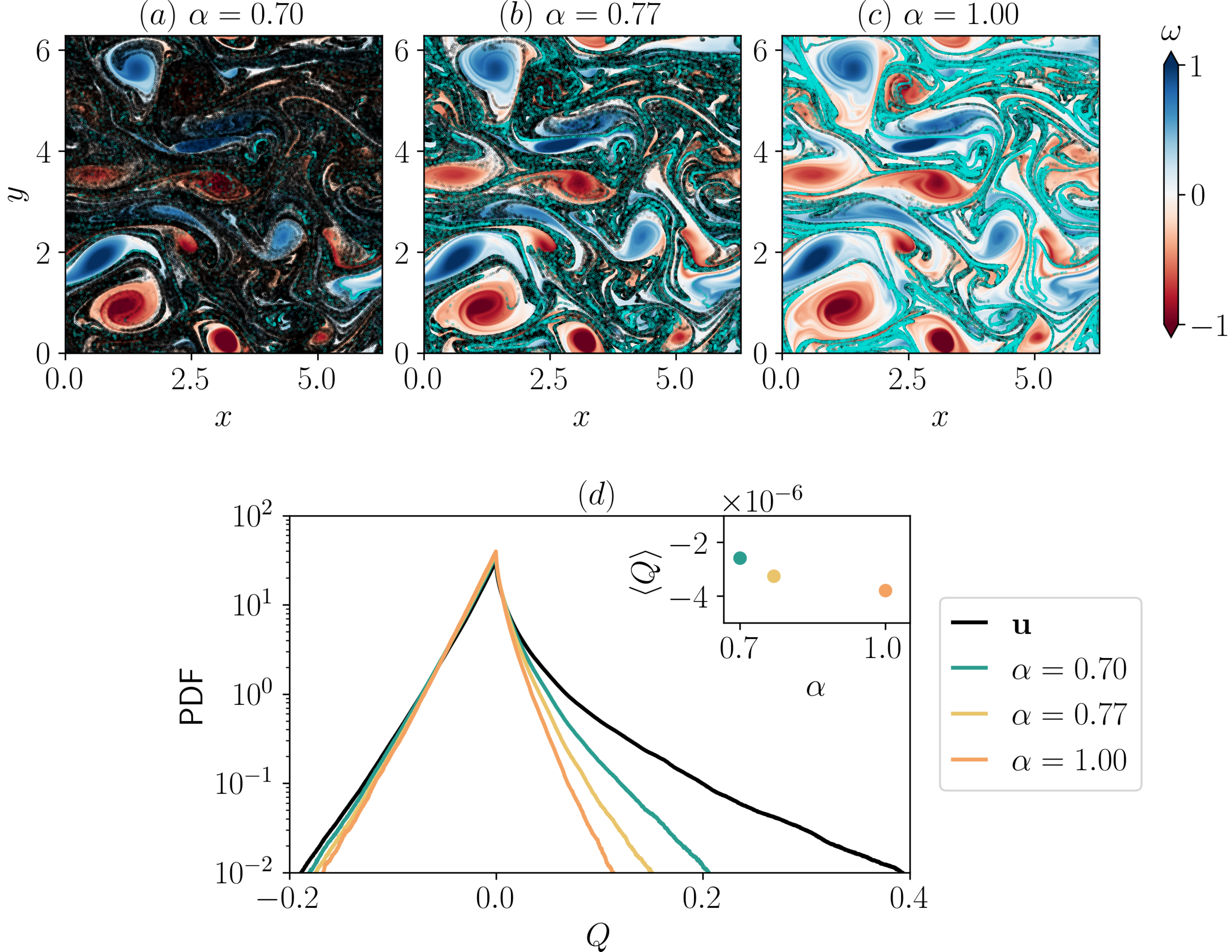}
\caption{(a-c) Snapshots of the vorticity field $\omega$ overlaid with caustics \cs particles in cyan, and the remaining ones in black at $t = 333 \ St$. All particles centrifuge out of large vortices and live in strain regions. As expected from prior studies \cite{balkovsky2001,calzavarini2008a, fiabane2012,karchniwy2019, petersen2019, motoori2023}, lowering the particle density weakens centrifugation. (d) PDF of the time-averaged $Q$ experienced by the particles (in color), as well as that of the background field (in black). Progressive emptying of high vorticity regions with increasing particle density is seen. }
\label{pstruc}
\end{figure}

Figure \ref{pstruc}(a-c) shows the vorticity snapshot of the background flow with inertial particles overlaid, for $\alpha=0.7, \ 0.77$ and $1.0$. Denser particles centrifuge out of vortices more readily than their lighter counterparts, as in previous studies \cite{balkovsky2001, calzavarini2008, calzavarini2008a, fiabane2012,karchniwy2019, petersen2019, motoori2023}, since lighter particles' centrifugation is hindered by their added mass \cite{balkovsky2001}. Denser particles show higher clustering levels in strain regions.  \cs particles (in cyan), increase in number with increasing particle density, and live alongside other particles in the strain region.  \fref {pstruc}(d) compares the distribution of $Q$ in the background field to those sampled by \emph {all} particles. As we know, extreme vortices are more common than extreme strain in 2D turbulence. The inset of \fref{pstruc}(d) shows that the mean $Q$ experienced by particles is a small negative number for particles, whereas this quantity is zero for the background flow. Interestingly, the leftmost part of \fref{pstruc}(d) shows that denser particles are infrequent in visiting regions of extreme strain as compared to their lower-density counterparts. We think this is because high strain is associated with rapid acceleration and deceleration in the flow, and high-density particles are less able to adapt to these changes due to their inertia.

\subsubsection{Behavior of caustics or \cs particles}
\begin{figure}
  \centering
  \includegraphics[width = \linewidth]{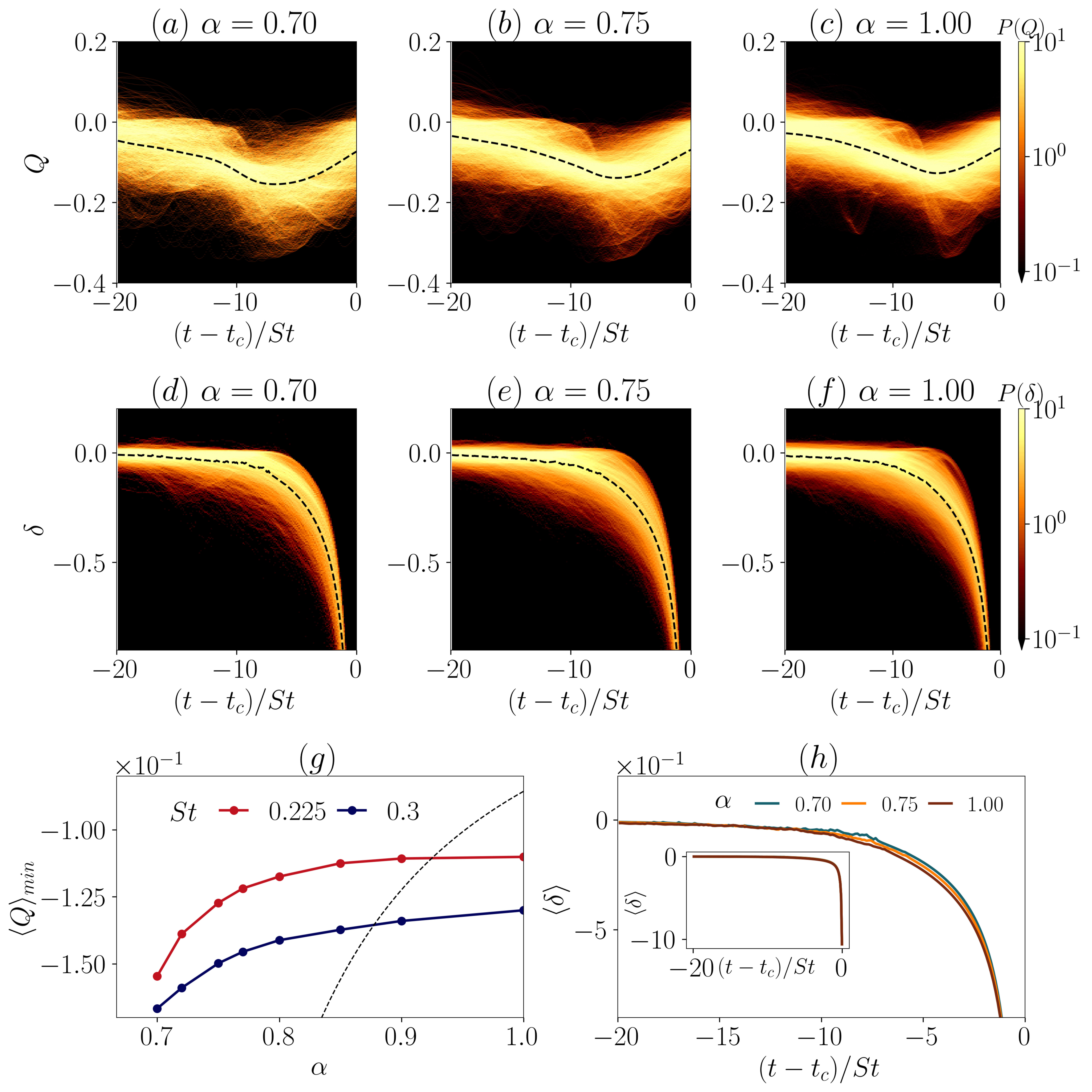}
  \caption{(a-c) PDF of the $Q$ explored by C particles of different density parameter $\alpha$. (d-f) the corresponding PDF of $\delta = \Tr(\mb{Z})$.  (g) Minimum of the average $Q$ explored by particles on their way to caustics as a function of  $\alpha$. (h) The mean of $\delta$ shows similar behaviour across density ratios with an inflection to large negative values around the time of minimum $Q$.}
  \label{fig: Q-timeseries-2d}
\end{figure}

{In \fref{fig: Q-timeseries-2d}, we shift the trajectories of each of the caustics particles by their caustics times $t_c$ to show a universal behavior leading upto caustics. As the probability density functions (PDFs) of Q in panels (a-c) show, each of the caustics particles experiences a maximum strain or minimum $Q = Q_m$ at time $t = t_m$. The black dashed lines show the mean of the Q-PDFs, $\mean{Q}$, with a minimum, $\mean{Q}_{min}$, occurring around $5 - 8$ Stokes times before caustics, denoted by $t = t_{min}$. Note that $\langle{Q})_{min}$ and $t_{min}$ are quantities averaged over caustics particle trajectories, while $Q_m$, $t_m$, and $t_c$ are different for different particles.   }

\fref{fig: Q-timeseries-2d}(g) shows $\mean{Q}_{min}$ against $\alpha$ for two Stokes numbers. Lower density particles have lower $\mean{Q}_{min}$, indicating that they need to go through regions of more intense strain than their denser counterparts to form caustics. This is also true in point-vortex flow, as seen in the figure, but how different this dependency is from $2$D turbulence is also highlighted.

Figure \ref{fig: Q-timeseries-2d}(d-f) shows the PDF of $\delta \equiv \Tr(\mb{Z})$ for \cs particles. The $\delta$ of individual particles was shifted in time by $t_c$ before calculating the PDF. Figure \ref {fig: Q-timeseries-2d}(h) shows $\mean{\delta}$, the mean of $\delta$ {over different particle trajectories as a function of shifted time $t-t_c$}, and the curves of different $\alpha$ practically overlap. Around $t=t_{min}$, $\delta$ undergoes an inflection point. Beyond this, the magnitude decreases sharply, resulting in caustics at finite time. {Each particle experiences a maximum strain around $t = t_m$ that inevitably} triggers caustics at late times. This result, known for infinitely dense particles \cite{meibohm2023,batge2023,zhang2025}, also holds across density ratios, and underlines the importance of large strain regions in triggering caustics. An interesting final observation is that particles exit high-strain regions before $t_c$ and caustics are actually formed in moderate strain regions. Thus, there is a universal mechanism of caustics formation across all density ratios.

\subsubsection{Behavior of survivor \su particles}\label{nc-dets}
Apart from the \cs and \es particles seen before in point-vortex flows, another behavior is seen in turbulent flow, displayed by survivor or `S' particles. These experience similar strain levels as the \cs particles, but never form caustics. We investigate them  below.

\citet{batge2023} found that not all infinitely-dense particles that visit regions of high strain form caustics. They derived a caustics criterion based on the eigenvalue magnitude of the fluid gradient and the time spent in large gradient regions. However, their analysis assumes that $\mb{Z}$ and $\mb{A}$ commute with each other, which is not true in general, due to the non-normality of $\mb{A}$ \cite{kronborg2023}. A similar analysis for finitely-dense particles requires that in addition to $\mb{A}$, $D\mb{A}/Dt$ commute with $\mb{Z}$ as well. This is not a fair assumption, and so a straightforward extension of \cite{batge2023} for finitely-dense particles is not possible. We show below that the time spent in strain regions is not the fundamental driver of the different dynamics of C and S particles. Rather, it is the time spent in extensional strain by the latter, which endows them with different pathline curvature and velocity, and the time spent in strain is a consequence of this.

\begin{figure}
\centering
\includegraphics[width =\linewidth]{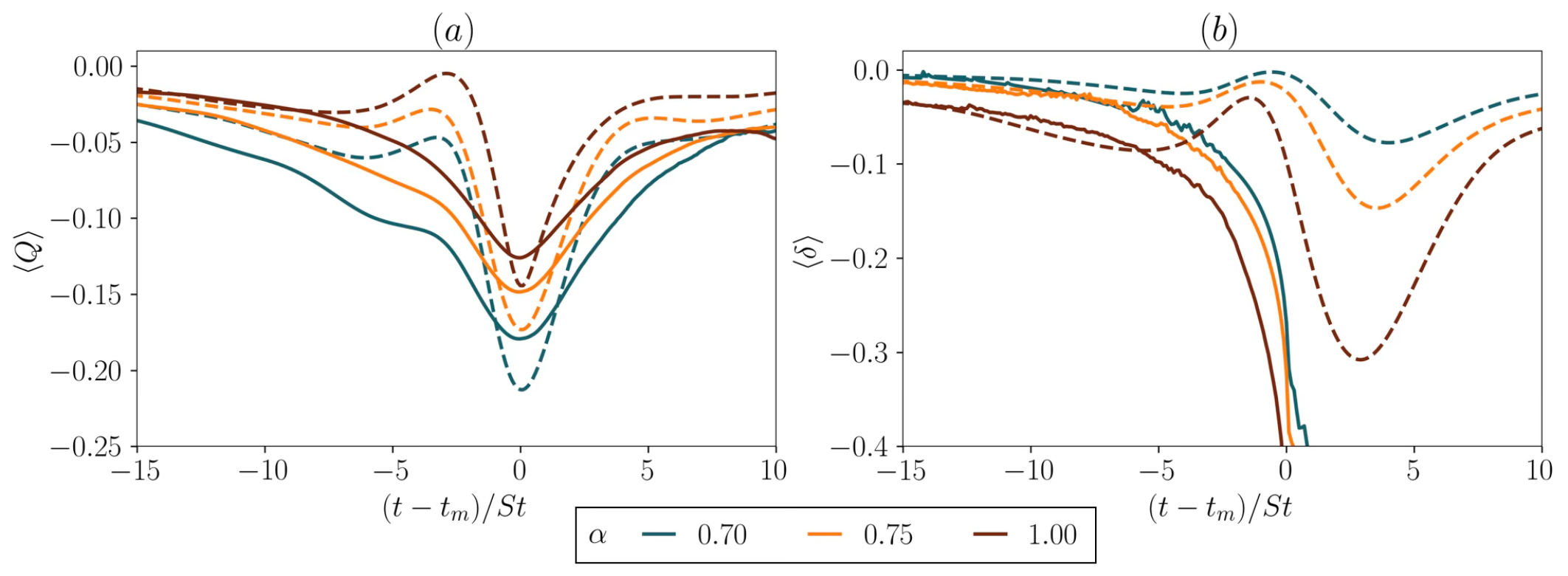}
\caption{Comparing (a) $\mean{Q}$ and (b) $\langle \delta \rangle$ of caustics \cs (solid lines) and survivor \su (dashed lines) particles around the time they encounter a minimum in $Q$.}
\label{c_nc_Q_trz}
\end{figure}

 To do this, we now shift the time of individual particles by their respective $t_{m}$ and define {$\mean{Q_m}_c$} as the mean of $Q_m$ experienced by \cs particles. Note that the time-shift makes {$\mean{Q_m}_c$} different from $\mean{Q}_{min}$. We define survivor \su particles as those that sample {$Q \le \mean{Q_m}_c$}, but do not form caustics. {After crossing $Q = \mean{Q_m}_c$, each of the survivor particles also experience a minimum  in $Q$ at time $t=t_m$. Accordingly, we shift their trajectory by $t_m$, and show the $\mean{Q}$ for both \cs and \su particles} as a function of the shifted time in \fref{c_nc_Q_trz}(a) for $\alpha = 0.7$, $0.75$ and $1.0$. After encountering the minimum in $Q$, all particles return to lower-strain regions of moderate $Q$. The broad trends are similar for \cs and \su particles, but \su particles explore larger negative $Q$ regions --- a consequence of our requirement that they visit a strain higher than {$\mean{Q_m}_c$} --- and also spend shorter times in them. We examine the differences in the history of \cs and \su particles to explain the difference in their ultimate fate. One clue is obtained from \fref{c_nc_Q_trz}(b), where we plot the mean of the $\delta = \Tr(\mb{Z})$ as a function of the shifted time $(t -t_m)/St$. Right before $t = t_m$, the $\mean{\delta}$ of \su particles rises to a maximum before plunging. This maximum is correlated with \su particles visiting low-strain regions (see the maxima in \fref{c_nc_Q_trz}(a)), which is not seen in the evolution of \cs particles. Thereafter, \su particles are subjected to a short duration of increasingly negative $\mean{\delta}$ post $t_m$. And unlike \cs particles, their $\delta$ recovers soon afterward, and they avoid caustics.  

To explain the similarity in $\mean{Q}$ and the dissimilarity in $\mean{\delta}$ of \cs and \su particles, we examine the strain and the vorticity of the background flow experienced by these particles, along with the curvature of the particle trajectories, and the alignment of the particle velocities with the local fluid strain directions. To do this, the velocity gradient tensor $\mb{A}$ of the incompressible flow $\vec{u}$ is written as 
\begin{align}
\begin{split}
    \mb{A} = \frac{St}{2}
    \begin{bmatrix}
      \sigma_n & \sigma_s + \omega \\
      \sigma_s - \omega & -\sigma_n
    \end{bmatrix}, 
\end{split}
\end{align}
where $\sigma_n = \partial_x u_x - \partial_y u_y, \sigma_s = \partial_x u_y + \partial_y u_x$, are the  normal and shear strains, respectively. The net strain given by $\sigma = \sqrt{\sigma_s^2 + \sigma_n^2}$. The eigenvalues of the strain tensor $\mb{S} = (\mb{A} + \mb{A}^T)/ 2$ are $\pm St~\sigma/2$, with the corresponding orthonormal eigenvectors 
\begin{align}
\label{epm}
\begin{split}
    \hat{e}_\pm = \frac{1}{\sqrt{(\sigma_n \pm \sigma)^2 + \sigma_s^2}}\left(\sigma_n \pm \sigma, \sigma_s\right).
\end{split}
\end{align} 
These eigenvectors $\epos$ and $\eneg$ represent the extensional and the compressional directions of the local strain of the background fluid, respectively. A particle in a strain region where the background fluid velocity is aligned with $\hat e_-$ experiences  compressional strain, whereas one aligned with $\epos$ experiences extensional strain. In addition, if the particle's velocity is close to that of the background fluid, which is true for small Stokes particles, an alignment with $\eneg$ represents the particle entering a saddle through compressional strain, and alignment with $\epos$ represents the particle exiting the saddle through extensional strain. The curvature of the particle's trajectory, defined as $|\dot{\hat{v}}(t)/|\vec{v}(t)||$, where $\hat{v}$ is the unit vector of $\vec{v}$, indicates whether the particle is moving in a straight line, or along a curved or zigzag path. The degree of alignment of the particle velocity with the compressional or extensional direction of the strain is measured by the alignment angle $\theta_v$, given by 
\begin{align}
\label{algn}
\begin{split}
\sin{\theta_v} = \hat{v}\cdot \epos, \ \ \cos{\theta_v} = \hat{v}\cdot \eneg \ .
\end{split}
\end{align} 
Thus, $\theta_v = 0,\pi$ represents a particle moving along the compressional direction of the strain, whereas $\theta_v = \pi/2,3\pi/2$ indicates that the particle is moving along the extensional strain direction. For small Stokes particles,  $\theta_v =\pi/2$ and $3\pi/2$ represent particles moving away from a local saddle point, whereas $0$ and $\pi$ represent motion towards it. Given the difference between these conditions, we do not expect complete top-down on left-right symmetry in any feature of $\theta_v$.

\begin{figure}
  \centering
  \includegraphics[width = 0.9\linewidth]{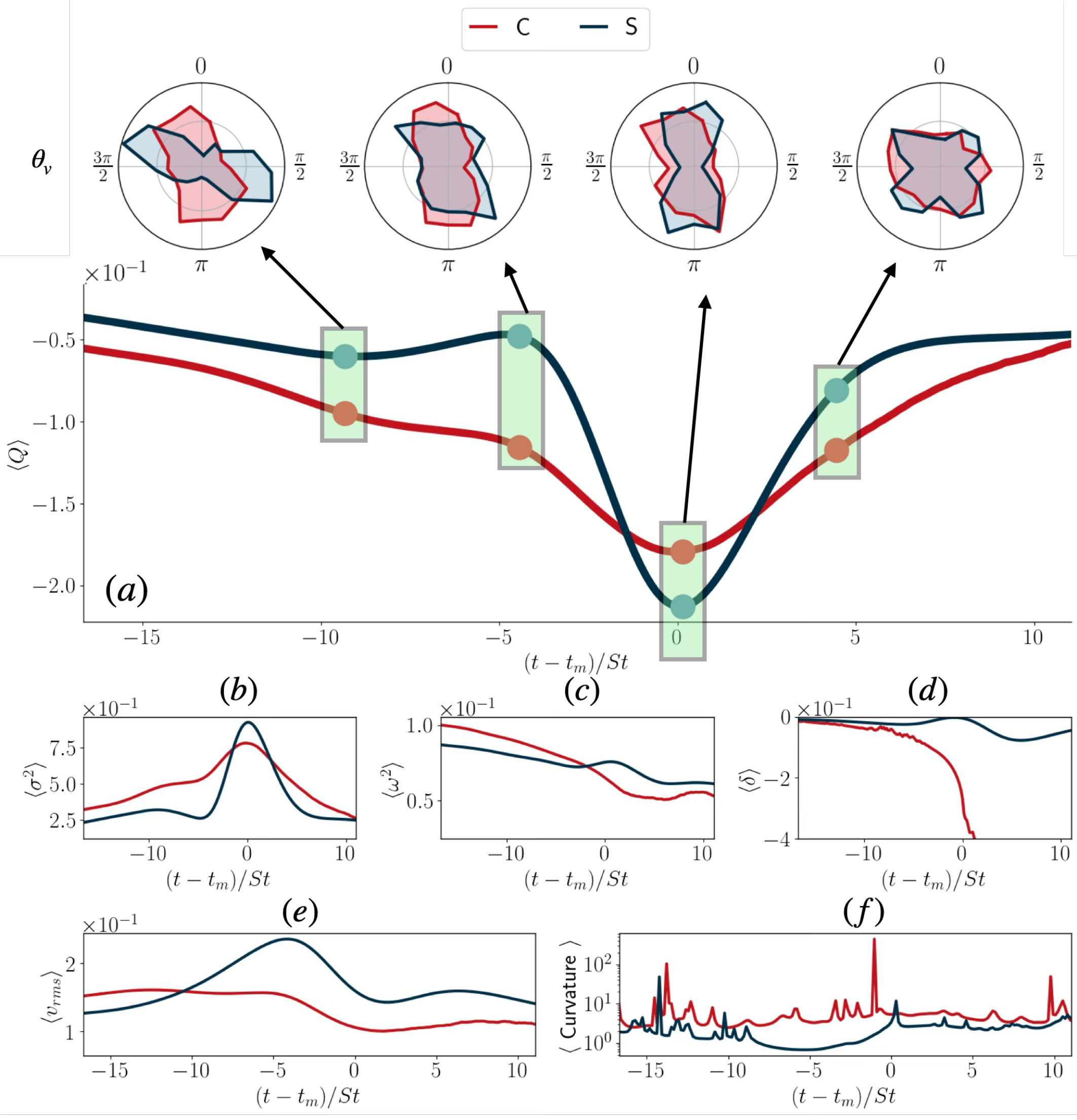}
  \caption{Comparing $\alpha = 0.7$ caustics and survivor particles using different metrics. Red is for \cs particles and blue for \su ones. (a) PDFs of the alignment angle $\theta_v$ are shown at times indicated by the green rectangles in the $\mean{Q}$ plot. (b,c) Average strain and vorticity experienced by \cs and \su particles during their evolution. (d) Average $\delta = \Tr(\mb{Z})$ for the \cs and \su particles. Despite similar $\mean{Q}$s, their $\mean{\delta}$ is significantly different. (e) RMS velocity of \su particles is higher than the \cs particles, and their curvature (f) is much lower for some time before caustics.}
  \label{caus_nc_sum}
\end{figure}

\fref{caus_nc_sum} highlights the differences between nearly-neutral \cs and \su particles in terms of the metrics discussed above. Panel (a) shows the PDF of $\theta_v$ at the times indicated in the $\mean{Q}$ plot. $\hat e_-(t)$ is calculated from \eref{epm} using the background strain at the location of the particle.
Due to the small $St$, $\vec{v}$ and $\vec{u}$ point in similar directions throughout.
A stark difference between \cs and \su particles is seen in the first panel, i.e., before the particles undergo sharply time-varying strains. With higher probability, \cs particles are better aligned with the compressional direction $\eneg$, whereas \su particles preferentially orient along the extensional direction of strain $\epos$, and experience a temporary reduction of strain, unlike \cs particles. Thereafter, both are catapulted into regions of high compressional strain (third panel) and finally (fourth panel) ejected into low-strain regions. The mean strain and vorticity magnitudes experienced by \cs and \su particles are provided in \fref{caus_nc_sum} (b) and (c) for completeness. Important differences are evident in \fref{caus_nc_sum} (e) and (f) which show respectively that just prior to entering the maximum in strain (minimum in $Q$), \su particles move faster than \cs counterparts and that their paths have significantly less curvature. Due to shooting straight and rapidly through the high strain region, \su particles spend less time there than \cs particles, and this is confirmed in the $\mean{Q}$ plot in (a). The culmination of these differences is manifested in a dramatic departure in behaviour of \cs and \su particles, as evidenced in \fref{caus_nc_sum} (d), where the former goes through caustics and the latter does not. Instead of a monotonic increase in negative $\delta$, \su particles actually show a reduction, going briefly to almost no divergence. 

\begin{figure}
\centering
\includegraphics[width = \linewidth]{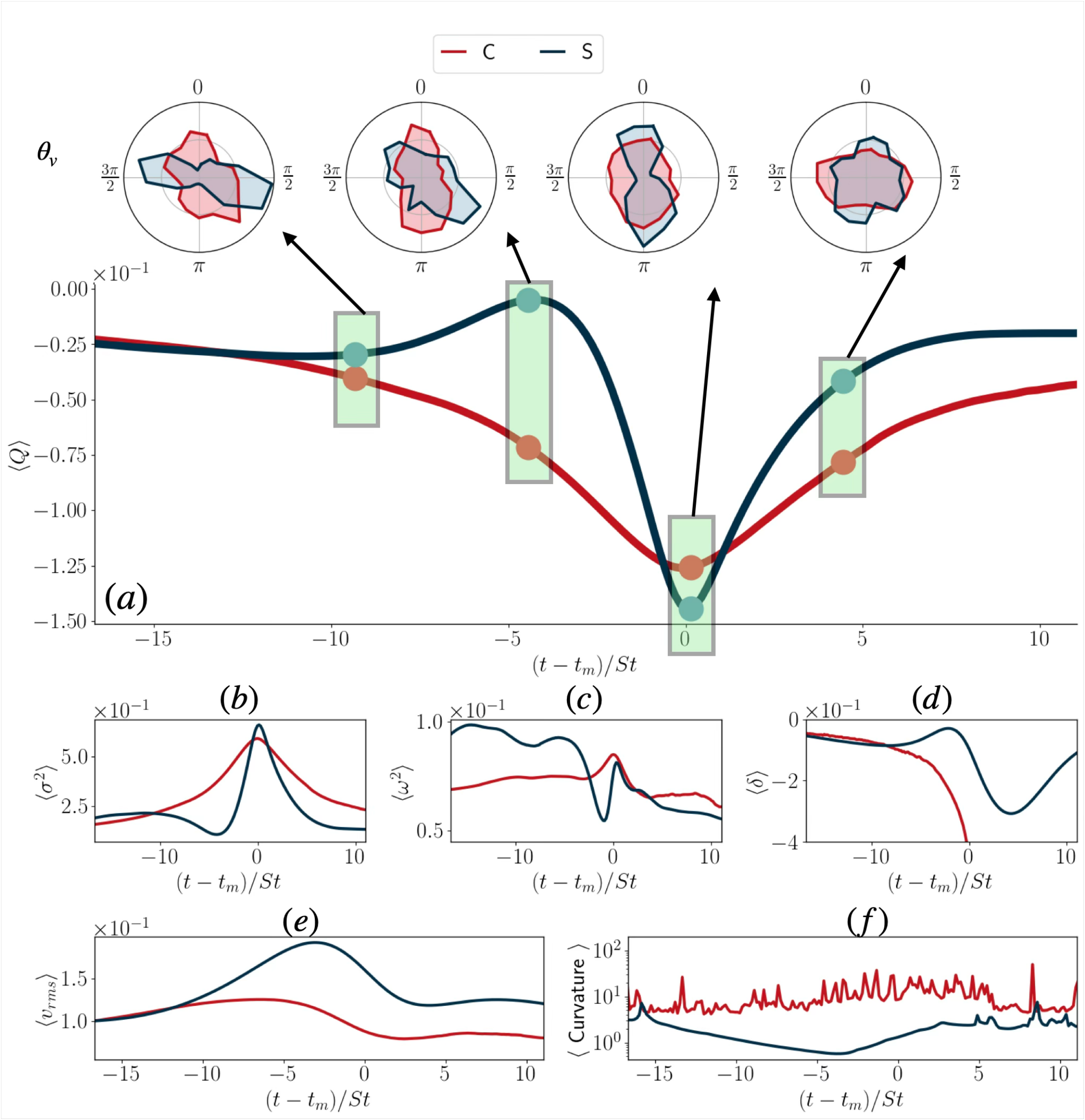}
\caption{Same as Fig. \ref{caus_nc_sum} but for inifinitely dense $\alpha = 1$ particles.} 
\label{caus_nc_sum_1}
\end{figure}

\begin{figure}
\begin{minipage}{0.475\linewidth}
\renewcommand{\arraystretch}{1.4}
\begin{ruledtabular}
\begin{tabular}{cccccc}
 & $\alpha$ & Caustics & Survivors & Avoiders \\
\hline
 & 0.70 & 0.0133 & 0.0662 & 0.9205 \\
 & 0.72 & 0.0306 & 0.1016 & 0.8678 \\
 & 0.75 & 0.0669 & 0.1233 & 0.8098 \\
 & 0.77 & 0.0906 & 0.1410 & 0.7684 \\
 & 0.80 & 0.1325 & 0.1431 & 0.7244 \\
 & 0.85 & 0.1970 & 0.1342 & 0.6688 \\
 & 0.90 & 0.2536 & 0.1193 & 0.6271 \\
 & 1.00 & 0.3099 & 0.1048 & 0.5854 \\
\end{tabular}
\end{ruledtabular}
\captionof{table}{Fraction of Caustics, Survivors, and Avoiders among $314000$ particles at time $t = 125 \tau_\eta$ with $St = 0.225$.}
\label{t1}
\end{minipage}\hfill
\begin{minipage}{0.475\linewidth}
\centering
\includegraphics[width = \linewidth]{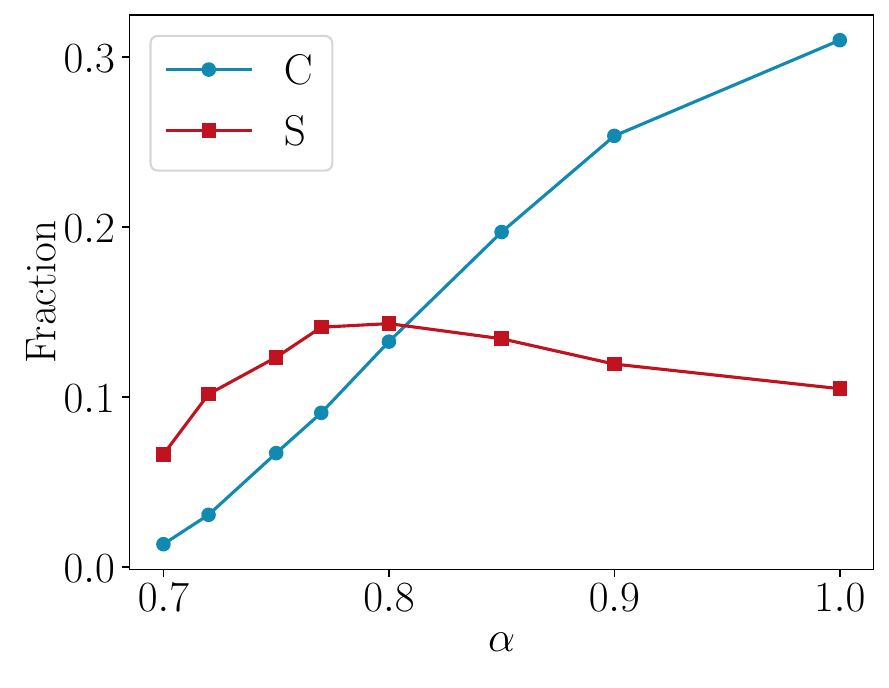}
\caption{Fractions of \cs and \su particles at different density ratios.}
\label{propensity}
\end{minipage}
\end{figure}

Therefore, we conclude that encountering a compressive strain with a moderate particle velocity is critical in staying in large strain regions long enough to form caustics. On encountering extensive strain regions, particles dash out with significant velocity. Even if they encounter a compressive strain on their way out, their velocity is too high to stay in that compressive strain long enough to form caustics. Thus, apart from the strain magnitude, the nature of the strain encountered by the particles, compressive or extensive, is critical in forming caustics as it controls the time particles spend in large strain regions. This conclusion also holds across density ratios, and Fig. \ref{caus_nc_sum_1} shows similar behavior of infinitely dense $\alpha = 1$ particles. {More such plots are included in the accompanying dataset \cite{chattopadhyay2026v2}, which validates our result for multiple Stokes numbers in the small Stokes regime, and across density ratios listed in table \ref{t1}. We have also examined our results by slightly altering the threshold $\mean{Q_m}_c$ for defining \su particles. While this changes the number of \su particles, the qualitative differences between \cs and \su particles remain preserved.
}

 Table \ref{t1} shows the relative numbers of particles in each category for different density ratios. Clearly, avoider \es particles are in the majority, and for near-neutral particles, visiting high strain (C and S) is a rare event. A non-monotonicity is noticed in \fref{propensity} in the relative fraction of C and S particles as $\alpha$ increases. This is because of two factors. The sum of the two goes up with $\alpha$ because the strain required for caustics is lower, as was seen in \fref{fig: Q-timeseries-2d}(g), and larger regions of the flow satisfy this. And at lower strain, the propensity to shoot out of extensional strain is lower, which makes survival difficult.

\section{Conclusion}\label{conc}
To summarize, we investigate caustics formation --- instances of extreme clustering --- of finitely-dense spherical point-like particles one-way coupled to a background flow, and compare them to their better-studied infinitely-dense counterparts. We restrict ourselves to particles denser than the fluid. Added mass and the acceleration in the background flow affect caustics formation of finitely-dense particles. 

We extend the caustics condition of \citet{meibohm2023} to the case of finitely dense frozen particles. The condition specifies regions of the background flow that can lead to caustics of particles with vanishing Stokes number. It involves invariants of the fluid gradient tensor, $Q$ and $R$, and the parameter $\alpha$ that captures the density ratio. $Q$ determines whether a region of the domain is vorticity dominated or strain dominated, while $R$ compares the relative strength of vortex stretching and strain amplification. We find that the relation says that for finitely-dense particles heavier than the background flow, regions of large compressive strain, characterized by regions of negative $Q$ and $R$, do not promote caustics. This contrasts with common intuition, where larger strain regions increase clustering of inertial particles denser than the fluid \cite{balkovsky2001}. {We also show that as the frozen-particle framework does not take into account the particles' residence time in high-strain regions, its application for finite Stokes particles is limited.}

\begin{figure}
  \centering
  \includegraphics[width = 0.5\linewidth]{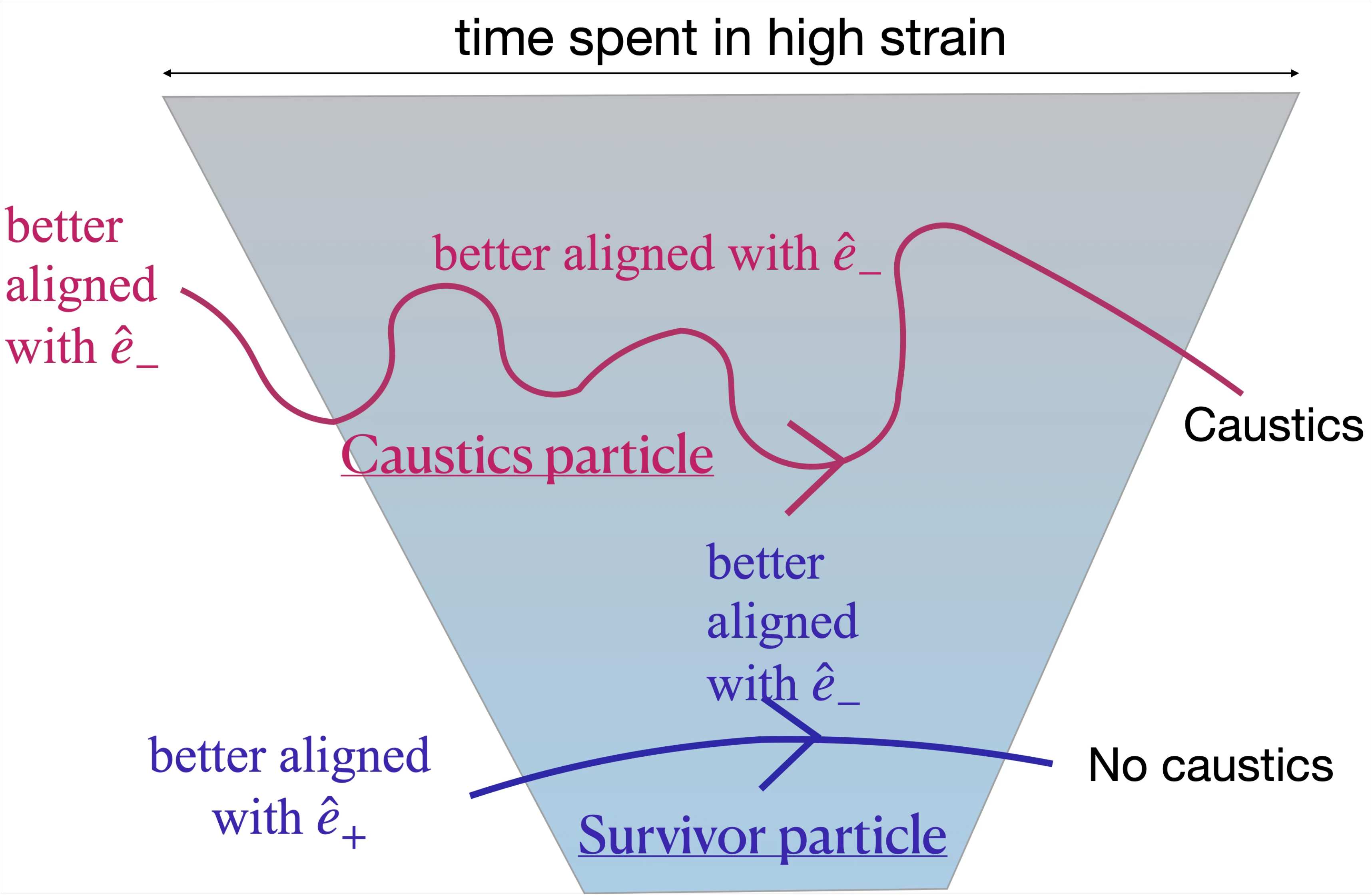}
  \caption{Schematic of differences between \cs and \su particles. The colored patch represents a region of high strain, and its horizontal extent the time spent by a particle within it. \su particles pass quickly through this region and follow relatively straight paths, whereas \cs particles spend longer and adopt high-curvature paths. The early alignment with the extensive and compressive eigendirections of strain is the cause for these differences. At later times, both C and S particle-velocities are randomly oriented.}
  \label{flchrt}
\end{figure}
We investigate caustics in two scenarios: point vortex flow and 2D turbulence. Particles that originate below a critical density-dependent distance from a point vortex form caustics. Scaling the time with $(3\alpha - 2)^{-1/2} $, a data-collapse is obtained in the near-vortex region. As the particle becomes lighter, the caustics region shrinks, i.e., the strain levels required increase, and the time for caustics formation gets longer, so it is harder for lighter particles to form caustics. The frozen-particle approach yielded a vastly different caustics region, and so, despite its analytical appeal, it is rather limited in reliability. Its predictions might hold for non-frozen particles with long residence times within the caustics region, but near a vortex, particles are typically centrifuged rapidly out of the vicinity.

In 2D turbulence too, only particles that go through high-strain regions can form caustics. As in point-vortex flow, caustics form after the particle has exited the high-strain region, and higher strain levels are needed for lighter particles. \su particles go through high strain but do not form caustics, and our explanation for this, holding for particles of any density, is depicted in a schematic in \fref{flchrt}. \su particles first encounter a region of extensional strain. They eject out of these regions and into compressional strain regions along low curvature and high velocity paths, transiting quickly through the high-strain region without forming caustics. In contrast, \cs particles approach regions of large compressive strain with smaller velocity and large curvature, stay there for a significant time, and form caustics. Thus, our analysis not only explains the reason behind the \citet{batge2023} findings for infinitely dense particles, but uncover the distinct history and mechanisms leading to the formation or the avoidance of caustics. 
In closing, we point out again that our findings are limited by our neglect of the Fax\'en and Basset-Boussinesq history forces on the particles. \citet{olivieri2014}, and \citet{daitche2015} find that particle clustering for near-neutrally buoyant particles diminishes due to history forces at O$(1)$ Stokes numbers, and we might expect a similar trend for caustics as well. Secondly, at higher $St$ and lower $\alpha$, $Re_p$ may not remain small, and we need to obtain suitably modified particle dynamics. {Particles with $\text{O}(1)$ Stokes are not as diverted by flow streamlines as smaller Stokes particles are. Thus, we expect them to be less affected by extensive or compressive straining regions and so our findings are valid for smaller Stokes numbers. }

{We note that our study applies to two-dimensional flows, and its extension to three dimensions is an important future direction. It is not straightforward to predict exactly how these findings will find relevance. There too, we may broadly expect that particles encountering compressional strain would undergo caustics, while those encountering extensional strain would survive. But three-dimensional flows have three strain directions, and isotropic turbulence mostly lives on the Vieillefosse line with positive $R$, preferring a plane (two directions) of extensive strain and an axis of larger compressive strain \cite{ashurst1987,meneveau2011}. Accordingly, infinitely dense particles show a minimum of $Q$ with a maximum of $R$ before caustics \cite{zhang2025}. Having a plane, rather than a line, of extensive strain might naively indicate lower caustics levels, but the magnitude of compressive strain is larger than both the extensive ones, and can lead to more caustics. }

{A separate issue is the lifetime of strong strain regions. Unlike in $2$D turbulence, where strong straining regions that connect the large coherent vortices are stable for long times, intense strains in $3$D turbulence are short-lived or intermittent \cite{buaria2022}. This also affects whether caustics particles face enough compressive strain to collide or survivor particles find enough extensive strain to escape. These questions can only be addressed by performing direct numerical simulations in $3$D turbulence, and differentiating time spent in each type of strain, which we aim to do in the future.} Additionally, non-spherical particles and higher particle number densities that force the flow and each other need to be studied in the future, and we hope our work will motivate such studies. 
\section{Acknowledgements}
We acknowledge support from the Department of Atomic Energy, Government of India, under projects No. RTI4013, and No. RTI4019. We thank Vijaykumar Krishnamurthy for access to {his GPU workstation. The later stages of the work were helped by the JC Bose Grant under the Anusandhan National Research Foundation of RG. Rajarshi thanks Divya Jagannathan, Saumav Kapoor, Tamoghna Ray, Jitendra Kethepalli, Mrinal Jyoti Powdel, and Siddhartha Mukherjee for helpful discussions. We have no competing interests, and Microsoft and GitHub copilots were used for code development and plot formatting. Lastly, we thank the two anonymous referees whose comments strengthened the paper.
\section{Data Availability}
The code to run the simulations is publicly available on \href{https://github.com/Rajarshi-prime/2D_Caustics}{GitHub}. The data and scripts that can remake the plots are available on Zenodo \cite{chattopadhyay2026v2}.

\appendix
\section{Derivation of caustics condition for frozen caustics} \label{caus-reg-drv}
To obtain \eref{caus_cond}, we extend the procedure outlined by \cite{meibohm2023} to the case of $\alpha \ne 1$. Defining the independent invariants of $\mb{A}$ and $\mb{Z}$ as 
\begin{align}
\begin{split}
  a = \Tr{\mb{A}} = 0; \quad Q = -\frac{1}{2}\Tr{\mb{A}^2}; \quad R = - \det{\mb{A}} = -\frac{1}{3}\Tr{\mb{A}^3};
\quad z_2 = \Tr{\mb{Z}^2}; \quad d =  \det{\mb{Z}},
\end{split}
\end{align}
we can write coupled algebraic equations for $\delta,z_2,\text{ and }d$ in terms of $Q$ and $R$ by taking different powers of \eref{root-eq}, taking their trace, and using Cayley-Hamilton theorem. Further, eliminating $z_2$ and $d$, and defining  $\delta \equiv \pm \sqrt{\zeta} - 3\alpha/2$, we get, after some algebra
\begin{align}
\begin{split}
   & \left(\zeta^2 + \frac{a_2 }{2}\zeta + a_0\right)^2 +  \zeta\left(a_1 - a_2a_0\right) = 0, \quad {\rm where} \label{pol4}\\
    a_0 &=  -\frac{3 \alpha ^4}{16}-(3 \alpha +1) \alpha ^2 Q+36 (\alpha -1) \alpha  R, \\
    a_1 &=-\frac{7 \alpha ^6}{16}-24 (7-3 \alpha ) (1-\alpha ) \alpha ^2 Q^2-\frac{1}{2} (21 \alpha -13) \alpha ^4 Q-288 (1-\alpha )^2 \alpha  Q R\\&\mathrel{\phantom{ =} }-1728 (1-\alpha )^3 R^2+4 (9 \alpha +7) \alpha ^3 R, \quad
    {\rm } \ a_2 = 3 \left[8 (1-\alpha ) Q-\alpha ^2\right]\ .
\end{split}
\end{align}
The condition for caustics reduces to not having any positive root of $\zeta$ in \eref{pol4}. This happens when $a_1 - a_2a_0 > 0$, which yields \eref{caus_cond}.

\bibliography{references} 
\end{document}